\documentclass[modern]{rnaastex}

\pdfoutput=1

\usepackage{microtype}
\usepackage{url}
\usepackage{amsmath}
\usepackage{amssymb}
\usepackage{natbib}
\usepackage{multirow}
\usepackage{graphicx}
\newcommand{\project}[1]{\textsf{#1}}

\newcommand{\ktwo}{\project{K2}}

\newcommand{\documentname}{\textsl{Note}}
\newcommand{\figureref}[1]{\ref{fig:#1}}
\newcommand{\Figure}[1]{Figure~\figureref{#1}}
\newcommand{\figurelabel}[1]{\label{fig:#1}}
\renewcommand{\eqref}[1]{\ref{eq:#1}}
\newcommand{\Eq}[1]{Equation~(\eqref{#1})}
\newcommand{\eq}[1]{\Eq{#1}}
\newcommand{\eqalt}[1]{Equation~\eqref{#1}}
\newcommand{\eqlabel}[1]{\label{eq:#1}}

\newcommand{\dd}{\ensuremath{ \mathrm{d}}}

\newcommand{\bvec}[1]{{\ensuremath{\boldsymbol{#1}}}}

\newcommand{\Normal}{\ensuremath{\mathcal{N}}}
\newcommand{\mA}{\ensuremath{\bvec{A}}}
\newcommand{\mC}{\ensuremath{\bvec{C}}}
\newcommand{\mS}{\ensuremath{\bvec{\Sigma}}}
\newcommand{\mL}{\ensuremath{\bvec{\Lambda}}}
\newcommand{\vw}{\ensuremath{\bvec{w}}}
\newcommand{\vy}{\ensuremath{\bvec{y}}}
\newcommand{\vt}{\ensuremath{\bvec{\theta}}}
\newcommand{\vm}{\ensuremath{\bvec{\mu}(\bvec{\theta})}}
\newcommand{\vre}{\ensuremath{\bvec{r}}}
\newcommand{\vh}{\ensuremath{\bvec{h}}}

\begin{document}\raggedbottom\sloppy\sloppypar\frenchspacing

\title{%
    Linear models for systematics and nuisances
}

\author[0000-0002-0296-3826]{Rodrigo Luger}
\affil{Department~of~Astronomy, University~of~Washington, Seattle, WA}

\author[0000-0002-9328-5652]{Daniel Foreman-Mackey}
\affil{Center for Computational Astrophysics, Flatiron Institute, New York, NY}

\author[0000-0003-2866-9403]{David W. Hogg}
\affil{Center for Computational Astrophysics, Flatiron Institute, New York, NY}
\affil{Center for Cosmology and Particle Physics, Department of Physics, New
York University, New York, NY}
\affil{Center for Data Science, New York University, New York, NY}
\affil{Max-Planck-Institut f\"ur Astronomie, Heidelberg, Germany}


\section{Introduction}

The target of many astronomical studies is the recovery of tiny astrophysical
signals living in a sea of uninteresting (but usually dominant) noise.
In many contexts (i.e., stellar time-series, or
high-contrast imaging, or stellar spectroscopy), there are structured
components in this noise caused by systematic
effects in the astronomical source, the atmosphere, the telescope, or
the detector.
More often than not, evaluation of the true physical model for these nuisances
is computationally intractable and dependent on too many (unknown) parameters
to allow rigorous probabilistic inference.

Sometimes, housekeeping data---and often the science data themselves---can
be used as predictors of the systematic noise.
Linear combinations of these predictors (or linear combinations of non-linear functions
of these predictors) are often used as
computationally tractable models that can capture the nuisances.
These models can be used to fit and subtract systematics prior to
investigation of the signals of interest, or they can be used in a
simultaneous fit of the systematics and the signals.
For our purposes, a \emph{linear model} for a column vector of data $\vy$
can be written in the form
\begin{align}
\eqlabel{linearmodel}
\bvec{y} = \vm + \mA \vw + \mathrm{noise}
\end{align}
where $\vm$ is the column vector expectation or mean model (the part of the model
that we care about), $\mA$ is a \emph{design matrix}, whose columns are basis
vectors (predictors) for the systematics,
and $\vw$ is the vector of weights or amplitudes, one
for each basis vector.
\newpage

Similar models have been used to describe the systematics in astrophysical time
series data \citep{Smith:2012, Wang:2016, Luger:2016}, galaxy or stellar spectra
\citep{Tsalmantza:2012, Ness:2015}, and imaging \citep{Fergus:2014,
Wang:2017}.
One issue with flexible data-driven models is their tendency to overfit and
reduce the astrophysical signal of interest.
This is generally tackled using a dimensionality reduction technique like
principal component analysis (PCA) or by applying strong priors or a
regularization to the weights vector $\vw$.

In this \documentname, we show that if a Gaussian prior is placed on the
weights $\vw$ of the linear components, the weights can be marginalized out
with an operation in pure linear algebra, which can (often) be made fast.
We illustrate this model by demonstrating the applicability of a linear model
for the non-linear systematics in \ktwo\ time-series data, where the dominant
noise source for many stars is spacecraft motion and variability.\\[0.125in]

\section{The problem}

Consider a dataset $\vy$ of $N$ measurements $y_i$ with covariance
matrix $\mC$.
In the common case of data collected with measurement error $\sigma_i$ on
individual data points but no correlation across measurements, $\mC$ is a
diagonal matrix with $\mC_{ij} = \sigma_{i}\delta_{ij}$, although in general
the off-diagonal elements capture the covariance between different
measurements. Given a linear model as in
\eq{linearmodel}, the probability of the data under the model is given by a
normal distribution with mean $\vm + \mA \vw$ and covariance $\mC$:
\begin{align}\eqlabel{likelihood}
p(\vy | \vt, \vw) &= \Normal(\vy; \vm + \mA \vw, \mC) \quad.
\end{align}
However, we are specifically not interested in the \emph{value} of $\vw$.
Instead, we will marginalize over it.
To perform this marginalization we must place a prior on $\vw$ that we will
assume to be Gaussian:
\begin{align}
p(\vw) &= \Normal(\vw; 0, \mL) \quad. \nonumber
\end{align}
With this prior and the likelihood in \eq{likelihood}, our goal is to marginalize
out the weights $\vw$; that is, we want to compute the marginalized
likelihood,
\begin{align}
\eqlabel{marglikeintegral}
p(\vy | \vt) &= \int_{-\infty}^{\infty} p(\vw)\,p(\vy | \vt, \vw) \dd\vw \quad.
\end{align}
In doing so, we would like to avoid explicitly solving for the weights $\vw$
while also avoiding the evaluation of numerical integrals.\\[0.125in]

\section{The solution}

As we show in the Appendix, the marginalized likelihood
(\eqalt{marglikeintegral}) may be expressed as:
\begin{align}
\eqlabel{p(y|t)normal}
p(\vy | \vt) = \Normal (\vy; \vm, \mC + \mA \mL \mA^\top) \quad.
\end{align}
This marginalized likelihood function can be numerically maximized to find the
maximum likelihood parameters $\theta^\star$, or it can be
multiplied by a prior $p(\theta)$ and used for posterior inference.
In either case, the evaluation of the model will include the effects of
marginalizing over $\bvec{w}$ in the linear model and any uncertainties in
those values will be propagated to the results.

It is often useful to compute the value of the linear model so that we can
``remove'' systematics from the data.
To derive this, we recognize that \eq{p(y|t)normal} is the likelihood of a
Gaussian Process.
This means that conditioned on the data and a choice of the parameters
$\bvec{\theta}$, the systematics will have a Gaussian distribution with mean
$\bvec{m}$ and covariance $\bvec{\Sigma}_\bvec{m}$ given by
\citep{Rasmussen:2006}
\begin{align}\eqlabel{pred}
\bvec{m} &= \mu(\bvec{\theta}) + \mA \mL \mA^\top  \left[\mC +
    \mA \mL \mA^\top\right]^{-1} \left[\bvec{y} - \mu(\bvec{\theta})\right]
    \nonumber\\
\bvec{\Sigma}_\bvec{m} &= \mA \mL \mA^\top - \mA \mL \mA^\top
    \left[\mC + \mA \mL \mA^\top\right]^{-1}
    \mA^\top \mL \mA \quad.
\end{align}\\

\section{The implications}

In the previous section, we presented an expression that can be used to
compute the likelihood function for a linear model marginalized over the
weights vector.
Linear models have been used throughout the astrophysics literature as
data-driven descriptions of complicated physical processes but, in some cases,
this analytic marginalization could be applied to improve performance---both
computational and statistical---of the models.
Linear models become more expressive as more basis components are added, but
they also become prone to overfitting.
A prior can be used to mitigate overfitting while maintaining the flexibility
of the model and the trick described in this \documentname\ can be used to
efficiently compute the likelihood marginalized over the many linear
parameters $\bvec{w}$.

\Figure{figure} shows an example where the marginalized likelihood function
described here is used to fit a data-driven systematics model to a light curve
from the \project{K2} mission.
The details of this model appear elsewhere \citep{Luger:2016, Luger:2017}, but
the basic idea is that this linear model can be used to describe the noise
introduced into the light curve by motion of the spacecraft's pointing.
This can be combined with a physical model of a transiting planet to
characterize the planet even when the signal is not visible in the raw data.

\begin{figure}[h!]
\begin{center}
\includegraphics[width=0.9\textwidth]{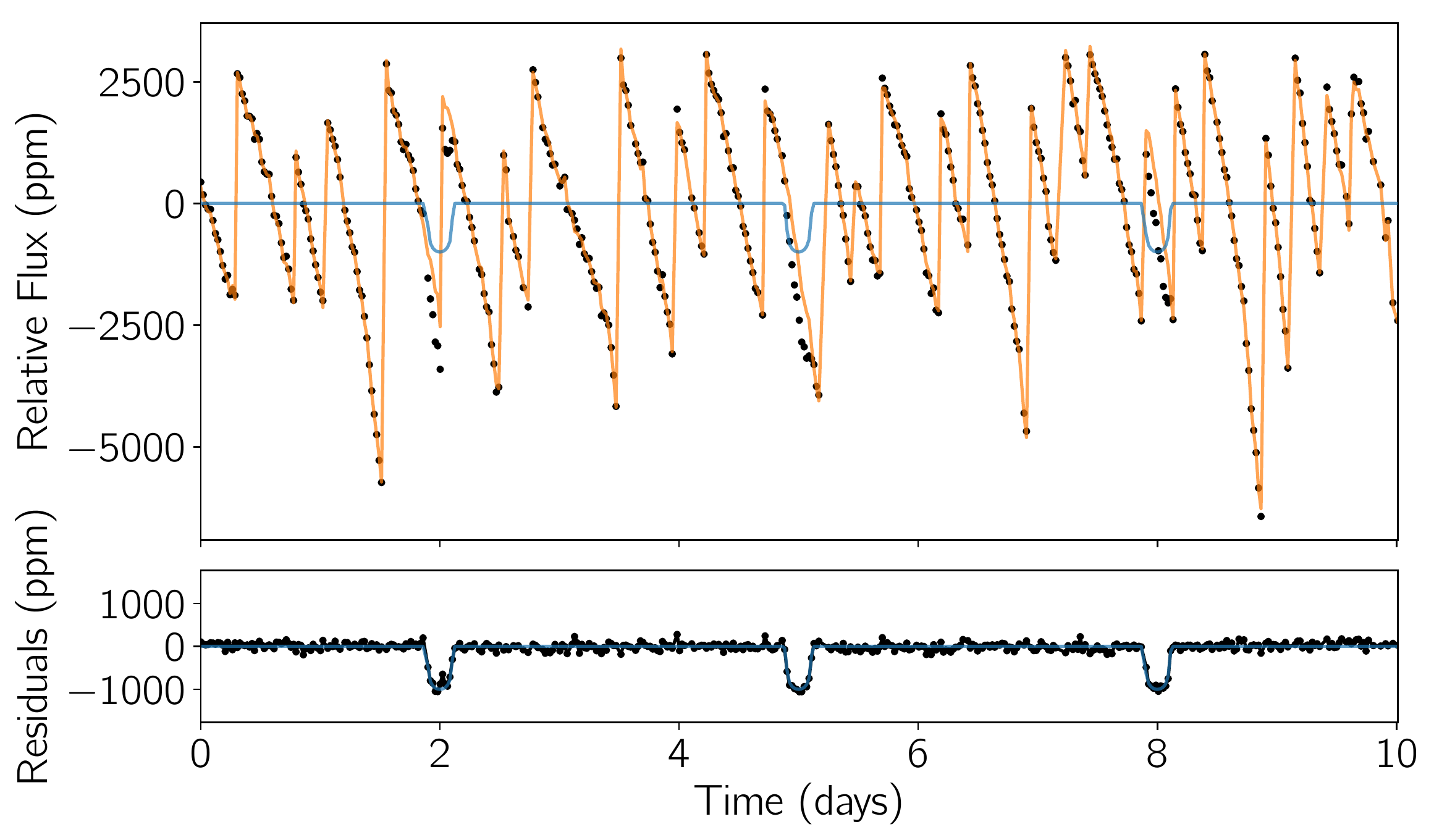}
\caption{%
    (top): The black points show the raw light curve for the \project{K2} target
    EPIC~204832142 multiplied by the time series for a simulated transiting
    planet.
    The simulated transit model is shown as a blue line.
    We fit the systematics using the linear model from the \project{everest}
    library \citep{Luger:2016, Luger:2017} and the prediction for the
    systematics model (\eqalt{pred}) is shown as an orange line.
    (bottom): The same data from the top panel with the systematics model
    subtracted.
    The transit model is plotted in blue.
\figurelabel{figure}}
\end{center}
\end{figure}

\acknowledgements
It is a pleasure to thank
  Patrick Cooper,
  Boris Leistedt,
  Bernhard Sch\"olkopf, and
  Dun Wang
for helping us understand all of this.

\bibliography{linear}

\begin{thebibliography}{}
\expandafter\ifx\csname natexlab\endcsname\relax\def\natexlab#1{#1}\fi
\providecommand{\url}[1]{\href{#1}{#1}}

\bibitem[{{Fergus} {et~al.}(2014){Fergus}, {Hogg}, {Oppenheimer}, {Brenner}, \&
  {Pueyo}}]{Fergus:2014}
{Fergus}, R., {et~al.} 2014, \apj, 794, 161

\bibitem[{Harville(1997)}]{Harville:1997}
Harville, D.~A. 1997, Matrix algebra from a statistician's perspective, Vol.~1
  (Springer)

\bibitem[{{Luger} {et~al.}(2016){Luger}, {Agol}, {Kruse}, {Barnes}, {Becker},
  {Foreman-Mackey}, \& {Deming}}]{Luger:2016}
{Luger}, R., {et~al.} 2016, \aj, 152, 100

\bibitem[{{Luger} {et~al.}(2017){Luger}, {Kruse}, {Foreman-Mackey}, {Agol}, \&
  {Saunders}}]{Luger:2017}
---. 2017, ArXiv e-prints, arXiv:1702.05488

\bibitem[{{Ness} {et~al.}(2015){Ness}, {Hogg}, {Rix}, {Ho}, \&
  {Zasowski}}]{Ness:2015}
{Ness}, M., {et~al.} 2015, \apj, 808, 16

\bibitem[{{Rasmussen} \& {Williams}(2006)}]{Rasmussen:2006}
{Rasmussen}, C.~E., \& {Williams}, K.~I. 2006, {Gaussian Processes for Machine
  Learning} ({MIT Press})

\bibitem[{{Smith} {et~al.}(2012){Smith}, {Stumpe}, {Van Cleve}, {Jenkins},
  {Barclay}, {Fanelli}, {Girouard}, {Kolodziejczak}, {McCauliff}, {Morris}, \&
  {Twicken}}]{Smith:2012}
{Smith}, J.~C., {et~al.} 2012, \pasp, 124, 1000

\bibitem[{{Tsalmantza} \& {Hogg}(2012)}]{Tsalmantza:2012}
{Tsalmantza}, P., \& {Hogg}, D.~W. 2012, \apj, 753, 122

\bibitem[{{Wang} {et~al.}(2016){Wang}, {Hogg}, {Foreman-Mackey}, \&
  {Sch{\"o}lkopf}}]{Wang:2016}
{Wang}, D., {et~al.} 2016, \pasp, 128, 094503

\bibitem[{{Wang} {et~al.}(2017){Wang}, {Hogg}, {Foreman-Mackey}, \&
  {Sch{\"o}lkopf}}]{Wang:2017}
---. 2017, ArXiv, arXiv:1710.02428

\bibitem[{Woodbury(1950)}]{Woodbury:1950}
Woodbury, M.~A. 1950, Memorandum report, 42, 336

\end{thebibliography}

\section{Appendix}

The marginalized likelihood may be expressed as follows:
\begin{align}
\eqlabel{integral}
p(\vy | \vt) &= \frac{1}{|2\pi\mL|^\frac{1}{2} |2\pi\mC|^\frac{1}{2}}
                \int_{-\infty}^{\infty} \exp \left[ -\frac{1}{2} z \right] \dd\vw
\end{align}
where
\begin{align}
\eqlabel{z}
z &= \vw^\top \mL^{-1} \vw + (\vre - \mA \vw)^\top \mC^{-1} (\vre - \mA \vw)
\end{align}
and $\vre = \vy - \vm$.
The integral is easier to evaluate if we
complete the square and write:
\begin{align}
\eqlabel{z_square}
z &= (\vw - \vh)^\top \mS^{-1} (\vw - \vh) + k \quad,
\end{align}
where, by comparison with \eq{z}, it can be shown that
\begin{align}
\mS^{-1} &= \mL^{-1} + \mA^\top \mC^{-1} \mA \\
\vh &= \bvec{\Sigma} \mA^\top \mC^{-1} \vre \\
k &= \vre^\top \left( \mC^{-1} - \mC^{-1} \mA \mS \mA^\top \mC^{-1} \right) \vre
    \quad.
\end{align}
We may thus write
%
\begin{align}
\eqlabel{p(y|t)ugly}
p(\vy | \vt) = \frac{1}{
                |2\pi\mL|^\frac{1}{2}
                |2\pi\mC|^\frac{1}{2}}
                \exp \left[ -\frac{k}{2} \right]
                \int_{-\infty}^{\infty} \exp
                \left[-\frac{1}{2}(\vw - \vh)^\top \mS^{-1} (\vw - \vh)
                \right] \dd\vw \quad.
\end{align}
The integral is that of a Gaussian, which evaluates to
$|2\pi\bvec{\Sigma}|^\frac{1}{2}$.
By the Matrix Determinant Lemma and the Woodbury Identity
\citep[for example,][]{Woodbury:1950, Harville:1997},
\begin{align}
|\mS| &= {|\mS^{-1}|}^{-1} = \frac{|\mL| |\mC|}{|\mC + \mA \mL \mA^\top|} \nonumber \\
k &= \vre^\top \left( \mC + \mA \mL \mA^\top \right)^{-1} \vre \quad.
\end{align}
Combining these results, the expression in \eq{p(y|t)ugly} simplifies to
\begin{align}
\eqlabel{p(y|t)exp}
p(\vy | \vt) &= \frac{1}{
                |2\pi(\mC {+} \mA \mL \mA^\top)|^\frac{1}{2}}
                \exp \left[ {-}\frac{1}{2} \big( \vy\,{-}\,\vm \big)^\top
                            (\mC {+} \mA \mL \mA^\top)^{-1}
                            \big( \vy\,{-}\,\vm \big)
                     \right].
\end{align}
This is a normal distribution with mean $\vm$ and covariance
$\mC + \mA \mL \mA^\top$:
\begin{align}
\eqlabel{p(y|t)normal_app}
p(\vy | \vt) &= \Normal (\vy; \vm, \mC + \mA \mL \mA^\top) \quad.
\end{align}

\end{document}